\documentstyle[aps,prl]{revtex}
\begin{document}

\draft
\title{Infrared Studies of a La$_{0.67}$Ca$_{0.33}$MnO$_3$
Single Crystal:\\ Optical Magnetoconductivity in a Half-Metallic
Ferromagnet}
\author{A. V. Boris$^1$, N. N. Kovaleva$^1$, A. V. Bazhenov$^1$,
P. J. M. van Bentum$^2$, Th. Rasing$^2$,\\ S-W. Cheong$^3$,
A. V. Samoilov$^4$ and N.-C. Yeh$^4$}
\address{$^1$Institute of Solid State Physics, Chernogolovka, Moscow
distr.,
142432, Russia}
\address{$^2$Research Institute for Materials and High Field Magnet
Laboratory,\\ University of Nijmegen, 6525 ED Nijmegen, The Netherlands}
\address{$^3$Bell Laboratories, Lucent Technologies, Murray Hill, NJ
07974,\\
and Department of Physics and Astronomy, Rutgers University, Piscataway, NJ
08855}
\address{$^4$Department of Physics, California Institute of Technology,
Pasadena, CA 91125}
\date{\today }
\maketitle

\begin{abstract}
The infrared reflectivity of a $\rm La_{0.67}Ca_{0.33}MnO_3$ single crystal
is studied over a broad range of temperatures (78-340 K), magnetic fields
(0-16 T), and wavenumbers (20-9000 cm$^{-1}$). The optical conductivity
gradually changes from a Drude-like behavior to a broad peak
feature near 5000 cm$^{-1}$ in the ferromagnetic state below the Curie
temperature $T_C=307\ K$. Various features of the optical conductivity bear
striking resemblance to recent theoretical predictions based on the
interplay
between the double exchange interaction and the Jahn-Teller electron-phonon
coupling. A large optical magnetoconductivity is observed near $T_C$.
\end{abstract}


Perovskite manganites of $\rm R_{(1-x)}A_{x}MnO_3$ ($\rm R$: trivalent
rare-earth ions, $\rm A$: divalent alkaline-earth ions, $0.2\leq
x\leq 0.5$) are
materials with interesting electric and magnetic properties which give
rise to novel phenomena such as the colossal negative magnetoresistance
(CMR). Many structural \cite{Radaelli,Louca}, magnetic
\cite{Ibarra,Shengelaya,Oseroff}, and transport \cite{Kaplan,Jaime,Zhao}
aspects of a ferromagnetic (FM) metal - paramagnetic (PM) insulator
instability, triggered by either temperature
or external magnetic field, can be explained in terms of the interplay of
the electron-phonon coupling and the double-exchange (DEX) transport
mechanism \cite{Millis1,Roder}. A spin polarized half-metallic behavior
exists in the
ferromagnetic state, which is believed to be essential for the
occurrence of CMR in the perovskite manganites
\cite{Pickett,Hwang,Okimoto,Wei}. However, details of the interplay between
the Jahn-Teller(JT) type electron-phonon coupling and the DEX mechanism
\cite{Millis1,Roder} are yet to be manifested experimentally.

This work aims at addressing this important issue by correlating the
infrared (IR) optical conductivity of a $\rm La_{0.67}Ca_{0.33}MnO_3$
single
crystal with its magnetic and electron transport properties. The
experimental approach involves measurements of the IR reflectivity spectra
of a $\rm La_{0.67}Ca_{0.33}MnO_3$ single crystal over a broad range of
temperatures (78-340 K), magnetic fields (0-16 T), and wavenumbers
(20-9000 cm$^{-1}$). In addition, dc magnetization and resistivity
measurements are performed on the same sample for comparison. The observed
temperature dependence of the optical conductivity bears striking
resemblance to the theoretical predictions \cite{Millis1,Roder}
of temperature-dependent competing effects between the JT coupling and
the DEX interaction in the ferromagnetic state. The frequency dependence of
the optical conductivity $\sigma (\omega)$
varies from the Drude-like behavior at low temperatures
to a broad peak feature in the mid-IR at high temperatures. These
characteristics of optical conductivity are shown to be determined by the
ratio of the electron-phonon coupling to the electron kinetic energy,
consistent with theory \cite{Millis1,Roder}.

The single crystal of $\rm La_{0.67}Ca_{0.33}MnO_3$ was
grown by a floating zone method. X-ray diffraction confirms that the sample
is single-phased with an orthorhombic structure close to cubic. The
temperature-dependent magnetization of this $\rm La_{0.67}Ca_{0.33}MnO_3$
single crystal reveals that the Curie temperature $T_C$ is at 307 K
($\pm$1 K), and the resistivity $\rho (T)$ shows a monotonic decrease with
decreasing temperature over the region investigated, with a significant
change in the slope at $T_C$.

Near-normal incidence reflectivities were measured with a combination of
a Fourier transform spectrometer, ''Bruker'' IFS 113v, and grating
monochromator, to cover the frequency range from 20 to 25000 cm$^{-1}$.
The far-IR reflectivity spectra were studied in the Faraday geometry
over 20 - 800 cm$^{-1}$ by using a
''Bruker'' IFS 113v spectrometer equipped with a superconducting magnet of
field strength up to 18.5 T. The reflectivities were
calibrated against a reference Au-mirror in the far-IR region, and against
a reference Al-mirror at higher frequencies.

The reflectivity spectra of the $\rm La_{0.67}Ca_{0.33}MnO_3$ single
crystal
for different temperatures from above $T_C$ down to 78 K and for
wavenumbers up to 9000 cm$^{-1}$ are shown in Fig.1, with more details
up to 800 cm$^{-1}$ given in the inset. The reflectivity spectra
above
$T_C$ reveal three distinct optical phonon features near 170, 350, and 580
cm$^{-1}$. These features are interpreted as the cubic perovskite
vibration bands of the $F_{1u}$ symmetry \cite{Kim,Boris}. As the
temperature
decreases from 340 K to 78 K, the reflectivity displays a gradual increase,
consistent with the occurrence of a metallic FM phase at low temperatures.
The optical phonon modes evident above 250 K become
screened and eventually disappear in the metallic phase.

To obtain the optical conductivity through the Kramers-Kronig analysis, we
used a Hagen-Rubens extrapolation for the low-frequency region.
Extrapolation towards high frequencies was made by using the reflectivity
data of $\rm La_{(1-x)}Sr_xMnO_3$ from 25000 cm$^{-1}$ to 80000 cm$^{-1}$
\cite{Okimoto}, because these data are nearly independent of the
temperature
and doping level. For energies higher than 80000 cm$^{-1}$, an
extrapolation
using the $\omega ^{-4}$ frequency dependence was used.
The real part of the optical conductivity $\sigma (\omega )$ for different
temperatures from above $T_C$ down to 78 K is shown in Fig. 2.
The Drude approximation
with a frequency-independent scattering rate is used to describe the data
in the low-frequency limit, and a sum of Lorentz oscillator functions is used
to account for the contribution from the transverse optical (TO) phonons
and for the mid-IR bands. We find that at low temperatures ($T \leq $ 150 K),
the $\sigma (\omega )$ spectrum exhibits a dominant Drude-like
contribution,
yielding a plasma frequency $\omega _p \approx 2.6\times 10^4$ cm$^{-1}$,
a scattering rate $\gamma \approx 5.5 \times 10^3$ cm$^{-1}$, and a
high-frequency dielectric constant $\varepsilon _{\infty } \approx $ 7.5
at 78 K, as shown by the dashed line in Fig.2.
In the context
of half-metallic ferromagnetism below $T_C$, these parameters provide an
estimate for the majority carrier density per Mn, $n$, corrected by the
effective mass, $m^{\ast }$, via the relation $n m_e / m^{\ast } =
(\omega _p^2 m_e)/(4\pi N_{Mn} e^2) \approx $ 0.40-0.45,
where $N_{Mn}$ is the number of Mn atoms per unit volume.
Assuming an effective carrier concentration $n\simeq $2/3 which is
comparable to the doping level, we obtain a reasonable effective band mass
$m^{\ast } \simeq $ 1.5-2 $m_e$ \cite{note}.

With increasing temperature, we find that the spectral weight of the
Drude component gradually decreases. At temperatures
above 250 K, the spectral weight is appreciably transferred to the broad
mid-IR band. As shown in Fig. 2, the optical conductivity at $T \geq $250 K
exhibits a broad peak on a continuum, extending from
the mid-IR region to the lowest frequencies. 
Well above $T_C$ at T=340 K the detailed dispersion analysis shows that the 
optical conductivity includes the small Drude part and broad mid-IR band 
around 5000 cm$^{-1}$ resulting from the intraband electronic 
excitations near the Fermi level, as presented by the dotted line in Fig.2. 
In addition, the mid-IR spectrum is completed with the broad band
peaking at approximately 11000 cm$^{-1}$, outside the range of the figure, 
(as shown by the light solid line in Fig.2), which appears in PM phase and 
may be treated as contribution from the interband excitations.
In the inset of Fig.2, we show
the temperature dependence of the total spectral weight of the Drude and
mid-IR contributions derived from the Drude-Lorentz analysis, expressed in
terms of the effective electron density per Mn site:
\begin{equation}
N_{eff} = \frac{2 m_e}{\pi e^2 N_{Mn}}\int (\sigma _{Drude}(\omega )+
\sigma _{mid-IR}(\omega ))d\omega.
\end{equation}
As shown in the inset of Fig.2, $N_{eff}$ is not a conserved
quantity: It decreases monotonically with increasing temperature in the
FM state, and then exhibits a sharp drop at the Curie temperature.

The experimental observation of the frequency-dependent optical
conductivity
may be understood in terms of a theoretical model that incorporates
both the DEX transport mechanism and an
electron-phonon coupling effect \cite{Millis1}. In a dynamical mean-field
approximation, it is assumed that the $d$-electron in the $e_g$ orbital of
the Mn$^{3+}$ ion is coupled to 3 $t_{2g}$ spins of Mn ions via the Hund's rule
exchange coupling, and to the classical oscillators of the lattice through
the Jahn-Teller effect \cite{Millis1}. It is found that $T_C$,
$\sigma (\omega ,T)$, and $\rho (T)$ are functions of an effective
coupling parameter $\lambda$, defined as the ratio of the energy gain from
localizing an electron at the Mn site to the electron kinetic energy, $t$.
That is, $\lambda \equiv E_{JT}/t$ and $E_{JT}=g^2/k$, where $g$ is the
electron-phonon coupling strength, $k$ the force constant of the classical
oscillators \cite{Millis1}.
Comparing our data with Ref. \cite{Millis1},
we find that the effective coupling parameter $\lambda $ is comparable to
unity, suggesting a moderate electron-phonon coupling in
$\rm La_{0.67}Ca_{0.33}MnO_3$. The broad peak in the mid-IR optical
conductivity can be attributed to transitions between the two
$e_g$-orbitals
of adjacent Mn sites, and is therefore comparable to
the Jahn-Teller splitting. Comparing our optical conductivity data (which
display a broad peak at $\sim$ 5000 cm$^{-1}$ near $T_C \approx 307$ K)
with theoretical calculations of Ref. \cite{Millis1}, we obtain a crude
estimate of $t \sim E_{JT} \sim 0.3$ eV for $\rm La_{0.67}Ca_{0.33}MnO_3$
single crystals.
In accordance with the theory, the temperature-dependent spectral
weight $N_{eff}$ is proportional to the kinetic energy of
mobile carriers.
In the FM state, the effective hopping amplitude
$t^{eff} (m)$ is dependent on the normalized magnetization
$m = M/M_s$, and may be approximated by the expression
$t^{eff} (m) \approx t q(m) \equiv t \sqrt{(1+m^2)/2}$ \cite{Millis2}.
Using a saturated magnetization $M_s \approx 3.9 \mu _B$ for
$\rm La_{0.67}Ca_{0.33}MnO_3$ and the
empirical magnetization data $M(T)$ of the sample, the resultant
temperature dependent $m$ and $q(m)$ are illustrated in the inset of Fig.2,
along with $N_{eff}$. This comparison of $N_{eff}$ and $q(m)$ shows that
the decrease in $N_{eff}$ with temperature cannot be entirely accounted for by
the temperature dependence of $t^{eff} (m)$ alone.
We may consider another temperature dependent factor for $N_{eff}$ due
to the JT-induced suppression of the DEX interaction. This factor,
estimated by the expression
$\exp \lbrack - E_{JT} k_B T / (\hbar \omega _{ph} )^2 \rbrack$
\cite{Kugel},
with $\omega _{ph}$ being a characteristic phonon frequency,
is independent of the magnetization, unlike $t^{eff} (m)$. Combining the
temperature dependence of both $q(m)$ and the
exponential suppression factor, we obtain a consistent description for the
spectral weight $N_{eff} (T)$ at temperatures below $T_C$, as shown by the
dashed line in the inset of Fig.2, with a reasonable fitting
parameter $\lbrack E_{JT}/(\hbar \omega _{ph} )^2 \rbrack \sim 1/(0.045 \pm
0.005 \rm eV)$.
Despite good agreement between our data and the theoretical analysis
of Millis {\it et al.} \cite{Millis1} at $T < 250 K$,
we note that $N_{eff}(T)$ begins to decrease faster with temperature
than the theoretical prediction as $T$ becomes closer to $T_C$.
A sharp drop of $N_{eff}$ at $T_C$ suggests that the FM metal - PM
insulator
transition in the CMR manganites is associated with a fundamental change in
the electronic configuration, which results in a significant shift in the
spectral weight, from intraband excitations near the Fermi level in the FM
state to effectively interband excitations in the PM state, the latter
involving an excitation energy of the order of the Hund's rule exchange
energy \cite{Okimoto}.

Fig.3 shows the real part of the reduced optical
conductivity at phonon frequencies, derived by subtracting the electronic
background at different temperatures. The TO phonon
spectra are represented by three sets of splitted bands and   
fitted by six Lorentzian lineshapes with maxima at 164, 178,
343, 370, 579, and 605 cm$^{-1}$. The vertical dashed lines in
Fig.3 show the temperature dependence of the frequencies for the main
TO phonon modes derived by the dispersion analysis.
No discernible changes in frequency are found in the TO phonon modes of the
single crystal except of redistribution of oscillator strength within 
the splitted bands with decreasing temperature.
This is in contrast to the report on polycrystalline
$\rm La_{0.7}Ca_{0.3}MnO_3$, where the fine structure of the phonon modes
could not be resolved \cite{Kim}. 
The observed redistribution of intensities  
between 579 cm$^{-1}$ and 605 cm$^{-1}$ in favour of the high-frequency one 
is in qualitative agreement with the pulsed neutron scattering data on powder 
samples of $\rm La_{1-x}Sr_xMnO_3,\ 0 \leq x \leq 0.4$ \cite{Louca}. 
As reported by Louca {\it et al.} the number of the MnO$_6$ Jahn-Teller 
distorted octahedra is decreased with decreasing temperature below $T_C$ 
resulting in an increase of the average number of the short Mn-O lengths. 
This is consistent with the observed temperature variation (Fig. 3)  
of the phonon feature near 579 cm$^{-1}$ related to the Mn-O bond length 
modulation \cite{Kim,Boris}. 

The magnetic field effect on the optical conductivity is studied by
measuring the far-IR reflectivity spectra under a wide range of magnetic
fields (0 -16 T).
A gradual increase in the reflectivity for the entire far-IR
frequency range is observed as the external magnetic field increases from
0 to 16 T and for all temperatures above $T_C$ (as shown in Fig.4(a)-(b)
for 307 K and 330 K). To compare the temperature and magnetic field effect
in detail, we consider representative spectra
shown in Fig.4 (c)-(d). The reference optical conductivity spectrum in
zero-field and for a constant temperature is presented by the lower solid
curve in each panel. The upper solid curve represents the
spectrum taken in the maximum field $H$= 16 T and at a temperature equal to
that of the reference spectrum. The dashed curve in each panel corresponds
to the optical conductivity spectrum at a temperature lower than that
of the reference spectrum, with the same low frequency limit as the upper
solid curve. First, we note that in the vicinity of $T_C$,
($i.e.$, at $T$ = 307 K, see Fig. 4(c)), the effect on enhancing
the optical conductivity due to an external magnetic field (16 T) is found
to be comparable to that due to decreasing the temperature (by $\Delta T$
= 17 K) in the entire far-IR frequency range.
We estimate that the corresponding dimensionless scaling factor
$k_{B} \Delta T/\mu _{B}H $ gradually changes with increasing temperature
from $\approx $ 1.6 at $T_C$ to $ \approx 2.5$ at $T = 330 K$,
as follows from Fig.4(c)-(d).
We also note the significant increase of $\sigma (\omega)$,
as illustrated in Figs.4(a)-(d), with the increasing magnetic field.
The magnetoconductivity, as large as about 50\% at 5 T in the vicinity of
the Curie temperature, is observed and may be referred to as an optical CMR effect.

In summary, we find that the optical conductivity spectra of a
$\rm La_{0.67}Ca_{0.33}MnO_{3}$ single crystal manifest a crossover from
the Drude-like to a hopping conductivity below the FM to PM phase transition.
The spectral features are in good qualitative agreement with recent
theoretical predictions based on the interplay between the double-exchange
mechanism and the Jahn-Teller type electron-phonon coupling.
The optical conductivity, and
the dc magnetization and resistivity data, are consistent with a
moderate effective electron-phonon coupling \cite{Millis1}.
A large optical magnetoconductivity at the phonon
frequencies is reported for the first time.

This research at the Institute of Solid State Physics is supported by
Russian State Program under Grant No.96031. Part of the work done at Caltech
is supported by NASA/OSS and the Packard Foundation.

\newpage
\centerline{FIGURE CAPTIONS}
\begin{description}

\item FIG.1 The reflectivity spectra of the $\rm La_{0.67}Ca_{0.33}MnO_3$
single crystal for different temperatures, ($T_C=307$ K).
Inset: more details in the far-IR region.

\item FIG.2  The real part of the optical conductivity
of the single crystal at different temperatures ($T_{C}=307$ K).
Inset: $N_{eff}$ (closed circles), normalized $dc$ magnetization
at $H=0.5 T$ ($m=M/M_{S}, M_{S}\simeq 3.9\mu _{B}$), and the
correspondent $q(m)$ (solid curves) vs. temperature $T$.
The suppression factor (see text) is illustrated by the dashed line.

\item FIG.3. Far-IR region of the real part of the reduced optical
conductivity, derived by subtracting the electronic background at
different temperatures: Successive curves are offset by
40 $\Omega ^{-1}cm^{-1}$ for clarity.

\item FIG.4 (a-b) The far-IR reflectivity spectra of the $\rm
La_{0.67}Ca_{0.3}MnO_{3}$ single crystal for different magnetic fields
at (a) $T$ = 307 K; and (b) $T$ = 330 K. (c-d) Comparison of the effect
of applying an external magnetic field (16 T) and that of reducing
the temperature on the far-IR optical conductivity in the single
crystal at (c) $T$ = 307 K; and (d) $T$ = 330 K.

\end{description}

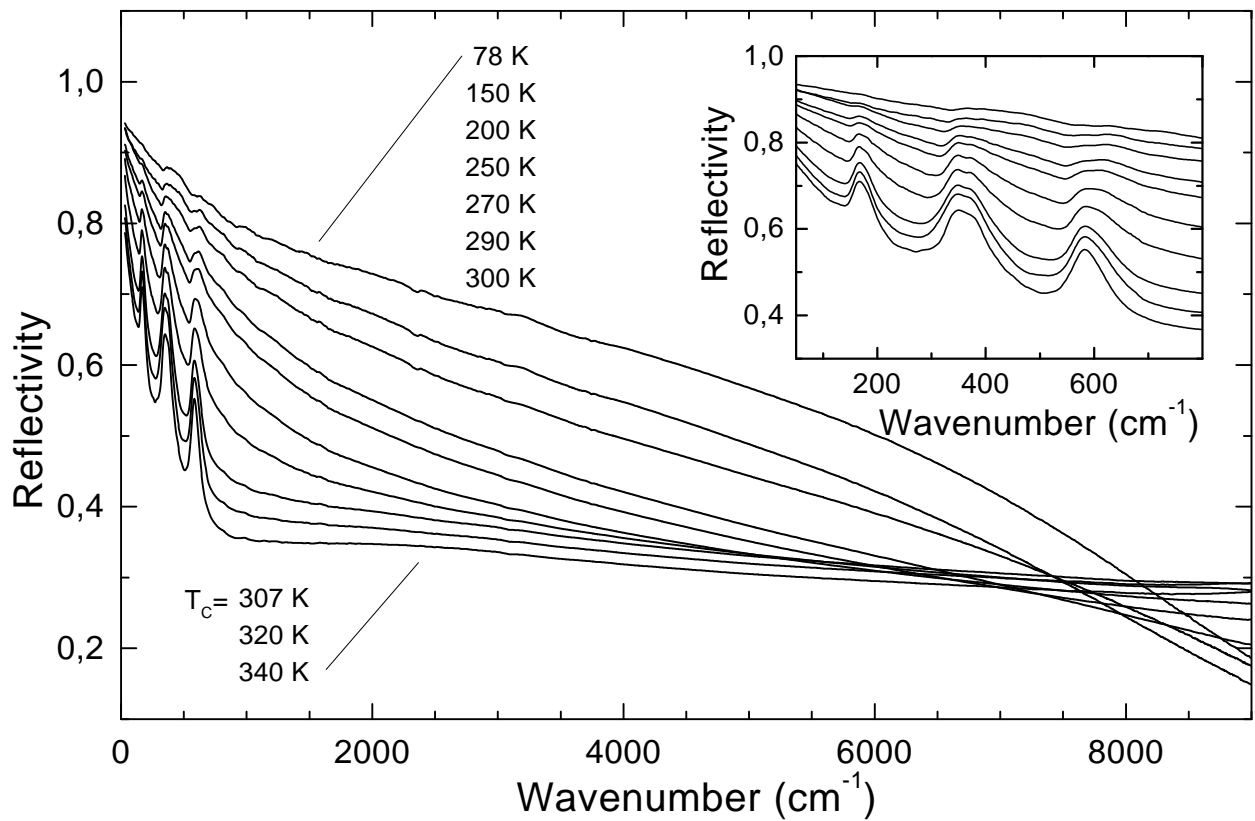
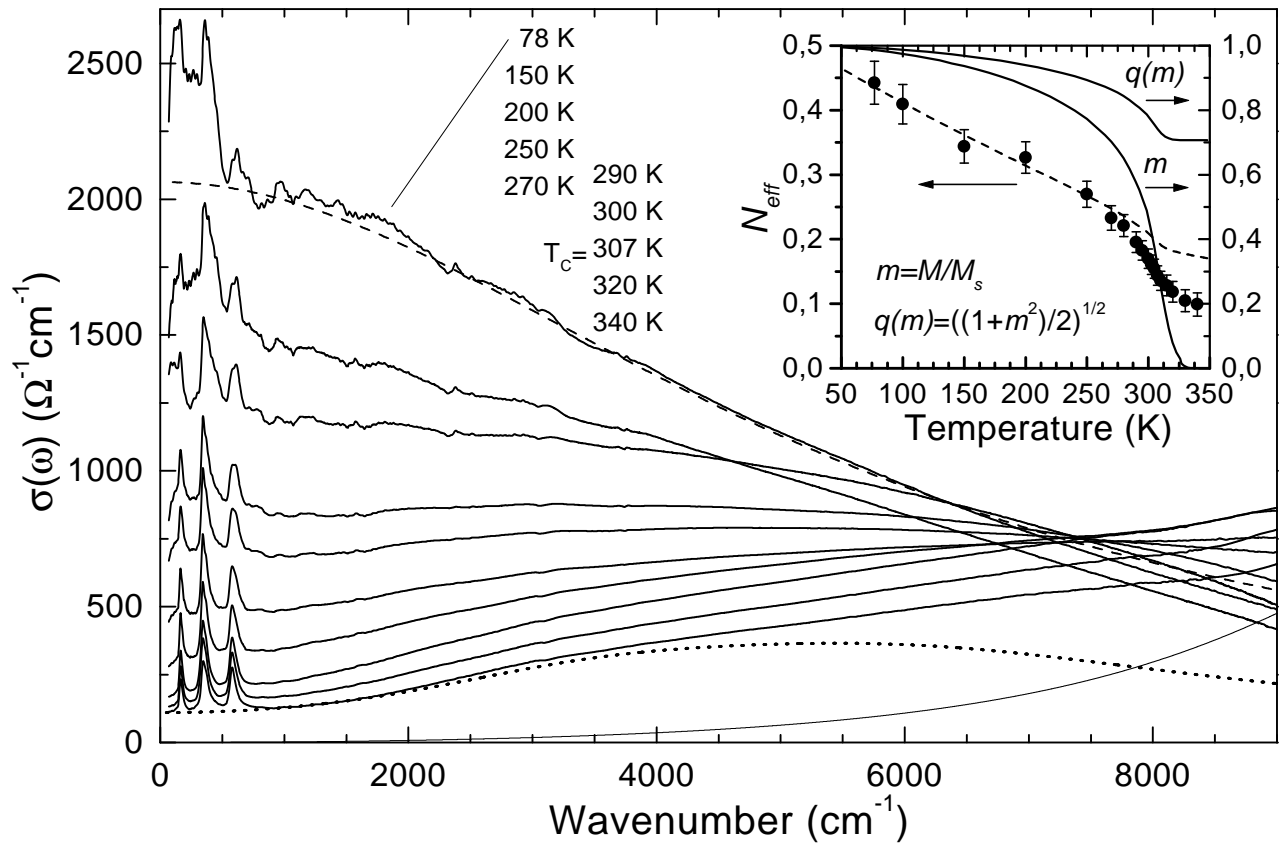
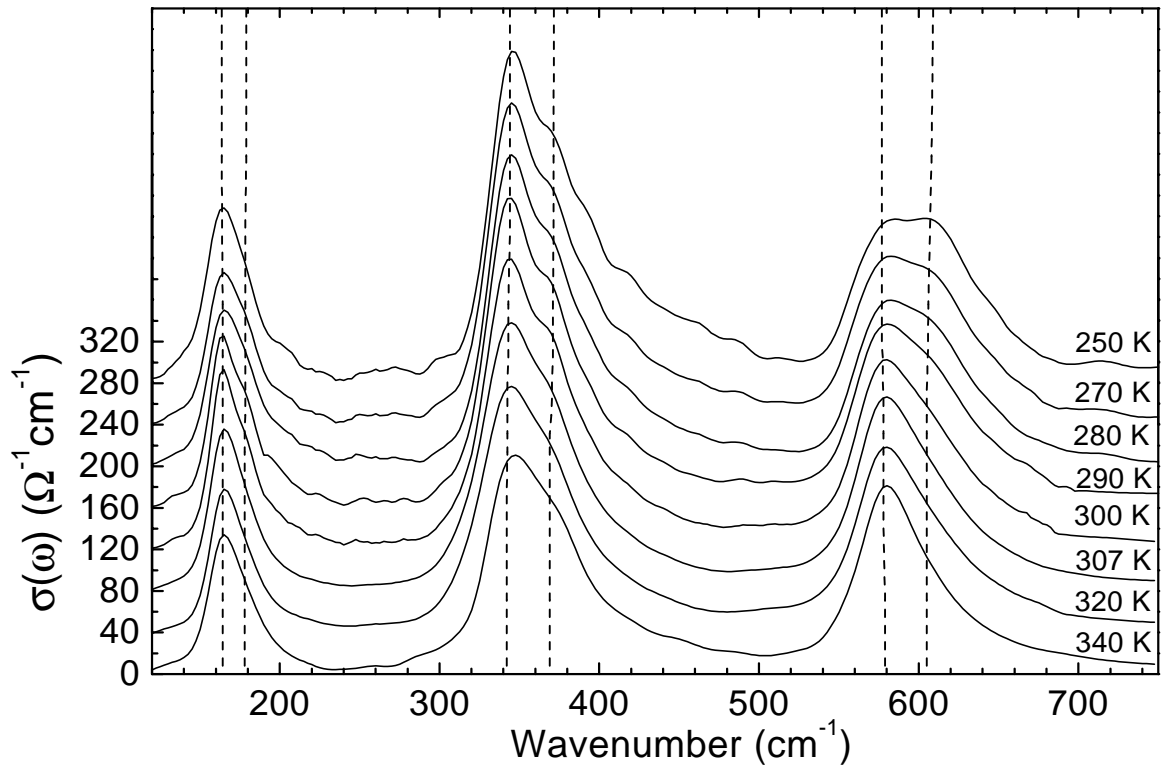
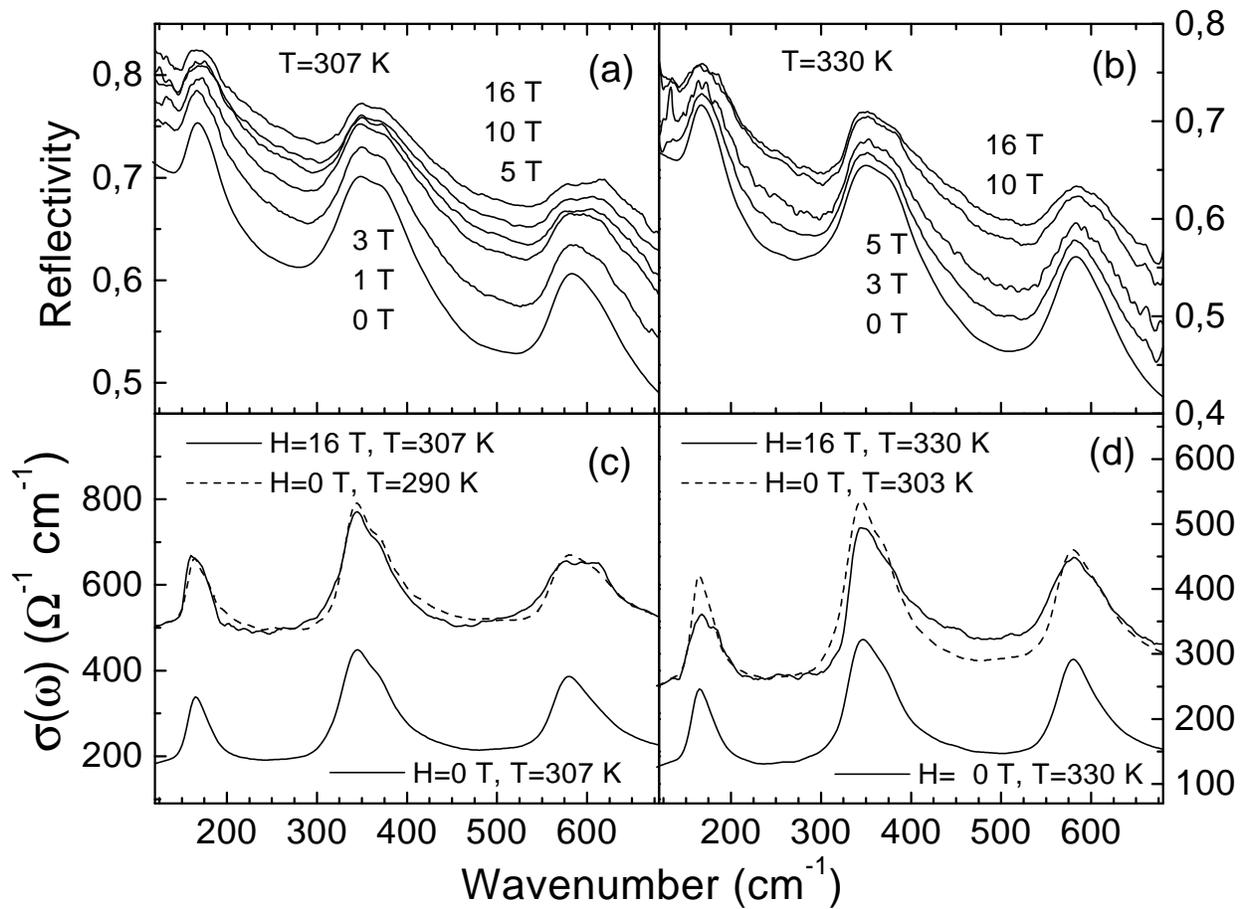
\end{document}